\documentclass[12pt,twoside]{article}

\usepackage{pslatex}	
\usepackage{amsfonts}
\usepackage{amsmath}
\usepackage{amsthm}
\usepackage{amscd}
\usepackage{cite}
\usepackage{epsf}
\usepackage{epsfig}
\usepackage{a4}

\def\beq{\begin{equation}}
\def\eeq{\end{equation}}
\def\bea{\begin{eqnarray}}
\def\eea{\end{eqnarray}}

\def\beann{\begin{eqnarray*}}
\def\eeann{\end{eqnarray*}}

  \let\g=\gamma \let\de=\delta
 \let\z=\zeta \let\h=\eta 

 \let\k=\kappa \let\la=\lambda 
 \let\x=\xi \let\p=\pi  \let\s=\sigma
\let\om=\omega 
\let\ph=\varphi  \let\PH=\Phi 
  
 \let\G=\Gamma \let\D=\Delta

\let\qd=\quad  

\def\epp{\, .}
\def\epc{\, ,}

\def\tst#1{{\textstyle #1}}

\theoremstyle{plain}

\newtheorem*{corollary*}{Corollary}

\theoremstyle{definition}

\def\2{\frac{1}{2}} \def\4{\frac{1}{4}}

\def\6{\partial}

\def\+{\dagger}

\def\<{\langle} \def\>{\rangle}

\let\auf=\uparrow

\renewcommand\i{{\rm i}}

\def\re{{\rm e}}

\def\tg{\, {\rm tg}\,}

\DeclareMathOperator{\sh}{sh}
\DeclareMathOperator{\ch}{ch}
\DeclareMathOperator{\tgh}{th}

\DeclareMathOperator{\cth}{cth}

\DeclareMathOperator{\arth}{arth}

\def\Re{{\rm Re\,}} \def\Im{{\rm Im\,}}

\def\res{{\rm res}}

\def\fa{\mathfrak{a}}
\def\faq{\overline{\mathfrak{a}}}

\renewcommand{\appendix}{%
   \renewcommand{\section}{
	\secdef\Appendix\sAppendix}%
   \setcounter{section}{0}%
   \renewcommand{\thesection}{\Alph{section}}%
   \renewcommand{\theequation}{\thesection.\arabic{equation}}%
}

\newcommand{\Appendix}[2][?]{%
     \refstepcounter{section}%
     \setcounter{equation}{0}%
     \addcontentsline{toc}{appendix}%
          {\protect\numberline{\appendixname~\thesection} #1}%
     \vspace{\baselineskip}%
     {\noindent\large\bfseries\appendixname: #2\par}%
     \sectionmark{#1}\vspace{\baselineskip}}

\newcommand{\sAppendix}[1]{%
     {\noindent\large\bfseries\appendixname\:: #1\par}%
     \sectionmark{#1}\vspace{\baselineskip}}

\renewcommand{\det}{\operatorname{det}}  

\begin{document}

\thispagestyle{empty}

\begin{center}

{\Large {\bf XX and Ising limits in integral formulae for
finite temperature correlation functions of the XXZ chain\\}}

\vspace{7mm}

{\large Frank G\"{o}hmann\footnote[2]{e-mail:
goehmann@physik.uni-wuppertal.de}
and Alexander Seel%
\footnote[3]{e-mail: seel@physik.uni-wuppertal.de}\\

\vspace{5mm}

Fachbereich C -- Physik, Bergische Universit\"at Wuppertal,\\
42097 Wuppertal, Germany\\}

\vspace{20mm}

{\large {\bf Abstract}}

\end{center}

\begin{list}{}{\addtolength{\rightmargin}{10mm}
               \addtolength{\topsep}{-5mm}}
\item
We consider a multiple integral representation for a one-parameter
generating function of the finite temperature $S^z$-$S^z$ correlation
functions of the antiferromagnetic spin-$1/2$ XXZ chain in the XX
limit and in the Ising limit. We show how in these limits the multiple
integrals reduce to single integrals, thereby reproducing known
results.
\\[2ex]
\end{list}

\clearpage

\section{Introduction}
The antiferromagnetic spin-$\2$ XXZ chain with Hamiltonian
\begin{equation} \label{xxzham}
     H = J \sum_{j=1}^L \Bigl( \s_{j-1}^x \s_j^x
            + \s_{j-1}^y \s_j^y + \D (\s_{j-1}^z \s_j^z - 1)
	      \Bigr)
\end{equation}
is a rare exception, even among the models solvable by Bethe ansatz
and connected to the Yang-Baxter equation, in that much is known
exactly about its correlation  functions. All the remarkable progress
achieved in the past decade originated in the derivation
\cite{JMMN92,JiMi96} of a so-called multiple integral representation
for the density matrix of the model. The next key works were
\cite{KMT99b}, where the results of \cite{JMMN92,JiMi96} were
rederived by means of the algebraic Bethe ansatz and generalized to
include a non-zero longitudinal magnetic field, then the article
\cite{BoKo01}, where some of the multiple integrals were actually
calculated for the first time, and \cite{BKS03}, where an ansatz for
the explicit form of a special density matrix element was formulated
(for the inhomogeneous and isotropic model).

The ideas developed in these works evolved further in various recent
articles. The ansatz of \cite{BKS03} for the inhomogeneous
generalization of the correlation functions was related to the
$q$-Knizhnik-Zamolodchikov equation, it was generalized to the
general density matrix element and finally proved (see
\cite{BJMST04a,BJMST04b} and the literature cited there). The
technique of \cite{BoKo01} was developed further, and a number of
concrete examples of short range correlation functions of the
XXZ chain was worked out, e.g. \cite{KSTS03,KSTS04}. Within the Bethe
ansatz based approach integral representations for correlation
functions other than the density matrix elements were developed
\cite{KMST02a}, the important problem of the summation of density
matrix elements into two-point functions was attacked successfully
\cite{KMST04cpp}, and multiple integral representations for
time-dependent correlation functions were suggested \cite{KMST04bpp}.

Based on a combination of the Bethe ansatz approach with quantum
transfer matrix techniques \cite{Suzuki85,InSu88,Kluemper92,Kluemper93}
the authors in collaboration with A. Kl\"umper derived multiple
integral representations for finite temperature correlation functions
\cite{GKS04a} and density matrix elements \cite{GKS04bpp,GKS05}.

This article is part of a programme of exploring the properties
of finite temperature correlation functions from their multiple
integral representations. So far the multiple integrals have proved to
be useful for deriving high temperature expansions for correlation
functions \cite{TsSh05} and have also been evaluated numerically
in the simplest non-trivial situation of three lattice sites
\cite{BoGo05}. Here we shall show how the multiple integrals reduce to
single integrals in the XX limit and in the Ising limit. As far
as the XX limit is concerned our discussion in many respects
parallels that for the zero temperature case \cite{KMST02c}.

The paper is organized as follows. In the next section we review
the multiple integral formula for the one-parameter generating
function \cite{IzKo84} of the $S^z$-$S^z$ correlation functions of the
XXZ infinite chain, that was obtained in \cite{GKS04a} and generalizes
a formula in \cite{KMST02a} to finite temperature. Then we treat the
XX limit and the Ising limit in two separate sections. We close with
a short summary.

\section{Integral representation of the generating function}
Starting point of our considerations in the following two sections
will be the multiple integral representation for a one-parameter
generating function of the $S^z$-$S^z$ correlation functions
derived in \cite{GKS04a}. This one-parameter generating function is
the thermal average
\begin{equation} \label{defgenfun}
     \PH (\ph|m) = \bigl\< \exp \{\ph \sum_{j=1}^m {e_j}_2^2 \}
                           \bigr\>_{T, h} \epc
\end{equation}
where $e_2^2 = \bigl( \begin{smallmatrix} 0 & 0 \\ 0 & 1
\end{smallmatrix} \bigr)$, and $h$ denotes the value of the
longitudinal magnetic field. The function $\PH (\ph|m)$ generates
a number of interesting correlation functions. We shall consider
the two-point function
\begin{equation} \label{genfunappl}
     \bigl\< S_1^z S_m^z \bigr\>_{T, h} = \tst{\frac{1}{4}}
          (2 D_m^2 \6_\ph^2 - 4 D_m \6_\ph + 1) 
                \PH (\ph|m) \Bigr|_{\ph = 0}
\end{equation}
and the so-called emptiness formation probability
\begin{equation} \label{defemptfp}
     P_\auf (m) = \lim_{\ph \rightarrow - \infty} \PH (\ph|m)
                = \bigl\< \prod_{j=1}^m {e_j}_1^1 \bigr\>_{T, h}
\end{equation}
which is the probability to find a ferromagnetic string of length
$m$ in the equilibrium state with temperature $T$ and magnetic field
$h$, and is a special matrix element of the density matrix. $D_m$ in
(\ref{genfunappl}) denotes the `lattice derivative' defined on any
complex sequence $(z_n)_{n \in {\mathbb N}}$ by $D_m z_m =
z_m - z_{m-1}$, and $e_1^1$ in (\ref{defemptfp}) is the matrix
$\bigl( \begin{smallmatrix} 1 & 0 \\ 0 & 0 \end{smallmatrix} \bigr)$.

We found in our previous works \cite{GKS04a,GKS05} that the same
auxiliary function $\fa$ that determines the free energy of the model
also contains all the information about the state of thermal
equilibrium that is necessary to express the correlation functions
as multiple integrals. As will become particularly clear in the
XX limit, the function $1/(1 + \fa)$ generalizes the Fermi
function to the interacting case. Since we expect the Fermi function
to appear naturally in all expressions for correlation functions,
we think that the formulation of the thermodynamics of integrable
models as developed in \cite{Kluemper92,Kluemper93} is particularly
useful in the present context.

\begin{figure}

\begin{center}

\epsfxsize \textwidth
\epsffile{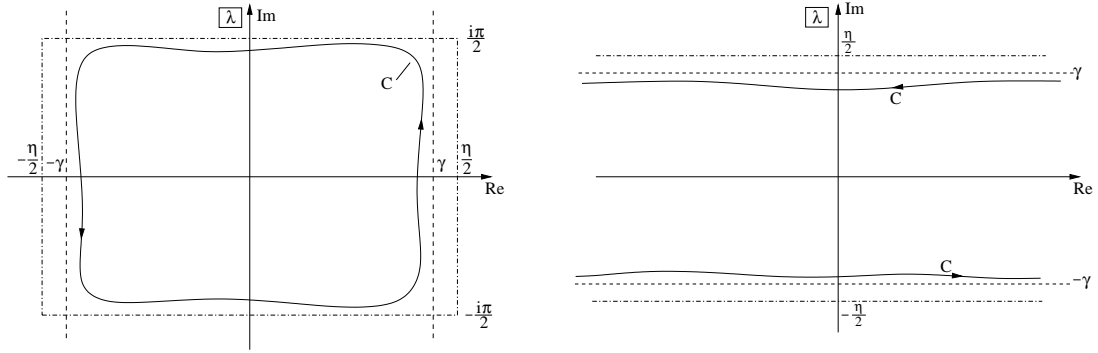}

\caption{\label{fig:cancon} The canonical contour ${\cal C}$ in the
off-critical regime $\D > 1$ (left panel) and in the critical regime
$|\D | < 1$ (right panel). For $\D > 1$ the contour is a rectangle
with sides at $\pm \i \frac{\p}{2}$ and $\pm \g$, where $\g$ is
slightly smaller than $\frac{\h}{2}$. For $|\D| < 1$ the contour
surrounds the real axis at a distance $|\g|$ slightly smaller than
$\frac{|\h|}{2}$ if $\D > 0$ and slightly smaller than $\frac{\p}{2}
- \frac{|\h|}{2}$ if $\D < 0 $.}
\end{center}
\end{figure}

The auxiliary function is the unique solution of the non-linear
integral equation
\begin{equation} \label{nlieh}
     \ln \fa (\la) = - \frac{h}{T}
                     - \frac{2J \sh^2 (\h)}{T \sh(\la) \sh(\la + \h)}
                     - \int_{\cal C} \frac{d \om}{2 \p \i} \,
	               \frac{\sh (2 \h) \ln (1 + \fa (\om))}
		            {\sh(\la - \om + \h)
			            \sh(\la - \om - \h)} \epp
\end{equation}
Here the integration contour ${\cal C}$ is a rectangle in the complex
plane. We have sketched it in figure \ref{fig:cancon}. Like the
inhomogeneity and the kernel in (\ref{nlieh}) it depends on the
parameter $\h$, which is the only essential parameter of the model
and parametrizes the anisotropy $\D$ in (\ref{xxzham}) as $\D =
\ch (\h)$.
Knowing the auxiliary function on $\cal C$ we can calculate the free
energy per lattice site according to the formula \cite{GKS04a}
\begin{equation} \label{freee}
     f (T,h) = - \frac{h}{2}
               - T \mspace{-5mu} \int_{\cal C} \frac{d \om}{2 \p \i} \,
                   \frac{\sh (\h) \ln (1 + \fa (\om))}
		        {\sh(\om) \sh(\om + \h)} \epp
\end{equation}

We can now state the main result of \cite{GKS04a}. The generating
function $\PH(\ph|m)$ of the correlation functions of the XXZ chain
in the thermodynamic limit, as defined in (\ref{defgenfun}), has the
following integral representation:
\begin{multline} \label{mirep}
     \mspace{-4mu} \PH (\ph|m)
     = \sum_{n=0}^m \frac{\re^{(m-n)\ph}}{(n!)^2}
	\biggl[ \prod_{j=1}^n \int_{\G}
	   \frac{d \z_j}{2 \p \i}
           \biggl( \frac{\sh(\z_j - \h)}{\sh(\z_j)}
	           \biggr)^{\mspace{-6mu} m} \mspace{-12mu}
	\int_{\cal C} \frac{d \om_j}{2 \p \i (1 + \fa (\om_j))}
        \biggl( \frac{\sh(\om_j)}{\sh(\om_j - \h)}
	              \biggr)^{\mspace{-6mu} m} \: \biggr] \\
        \biggl[ \prod_{j,k=1}^n
	     \frac{\sh(\om_j - \z_k - \h)}{\sh(\z_j - \z_k - \h)}
	          \biggr]
        \det_n M (\om_j,\z_k) \; \det_n G(\om_j,\z_k) \epc
\end{multline}
where $\fa (\la)$ is the auxiliary function (\ref{nlieh}), where
\begin{align}
     M (\om_j,\z_k) & = t(\z_k,\om_j)
        \prod_{\ell=1}^n \frac{\sh(\om_j - \z_\ell - \h)}
	                      {\sh(\om_j - \om_\ell - \h)}
	+ t(\om_j,\z_k) \re^\ph \prod_{\ell=1}^n
	  \frac{\sh(\om_j - \z_\ell + \h)}{\sh(\om_j - \om_\ell + \h)}
	  \epc \\
    t(\z, \om) & = \frac{\sh(\h)}{\sh(\z - \om) \sh(\z - \om + \h)}
          \epc
\end{align}
and where the function $G(\la,\z)$ is the solution of the linear
integral equation
\begin{equation} \label{repdefg}
     G(\la,\z) = t(\z,\la)
                  + \int_{\cal C} \frac{d \om}{2 \p \i (1 + \fa (\om))}
		    \, \frac{\sh (2 \h) G (\om,\z)}
		            {\sh(\la - \om + \h)\sh(\la - \om - \h)}
			    \epp
\end{equation}
The contour ${\cal C}$ in the latter equation is the same as in the
non-linear integral equation (\ref{nlieh}) (see figure
\ref{fig:cancon}). $\G$ is any simple closed contour which lies
inside ${\cal C}$ and encircles the origin counterclockwise.

An alternative expression for $\PH (\ph|m)$ which involves the
function $\faq = 1/\fa$ and which was also obtained in \cite{GKS04a}
is
\begin{multline} \label{mirepq}
     \mspace{-4mu} \PH (\ph|m)
     = \sum_{n=0}^m \frac{(-1)^n}{(n!)^2}
	\biggl[ \prod_{j=1}^n \int_{\G}
	   \frac{d \z_j}{2 \p \i}
           \biggl( \frac{\sh(\z_j + \h)}{\sh(\z_j)}
	           \biggr)^{\mspace{-6mu} m} \mspace{-8.mu}
	\int_{\cal C} \frac{d \om_j}{2 \p \i (1 + \faq (\om_j))}
        \biggl( \frac{\sh(\om_j)}{\sh(\om_j + \h)}
	              \biggr)^{\mspace{-6mu} m} \: \biggr] \\
        \biggl[ \prod_{j,k=1}^n
	     \frac{\sh(\om_j - \z_k + \h)}{\sh(\z_j - \z_k + \h)}
	          \biggr]
        \det_n M (\om_j,\z_k) \; \det_n G(\om_j,\z_k) \epp
\end{multline}
For the emptiness formation probability (\ref{defemptfp}) we obtained
several alternative expressions \cite{GKS04a,GKS04bpp}. In the present
context it will turn out to be convenient to work with the
inhomogeneous, symmetric version
\begin{multline} \label{pinintrep}
     P_\auf (m|\{\x\}) = \frac{1}{m!}     
	\biggl[ \prod_{j=1}^m
	        \int_{\cal C} \frac{d \om_j}{2 \p \i (1 + \fa (\om_j))}
	\biggr] \\[1ex]
	\biggl[ \prod_{j,k = 1}^m
	        \frac{\sh(\om_j - \x_k) \sh(\om_j - \x_k - \h)}
		     {\sh(\om_j - \om_k - \h)} \biggr]
        \frac{\det_m \, t(\x_k, \om_j) \det_m G(\om_j, \x_k)}
	     {\prod_{\substack{j,k = 1 \\ j \ne k}}^m
	      \sh(\x_j - \x_k)} \epc
\end{multline}
which follows from equation (111) of \cite{GKS04a} in the limit
$\ph \rightarrow - \infty$, and to perform the limit $\x_j \rightarrow
0$, $j = 1, \dots, m$, at a later stage.

Formulae for the magnetization are the simplest realizations of
the above integral representations, e.g.
\begin{equation}
     m(T,h) = \< S_j^z \>_{T, h}
            = \lim_{\x \rightarrow 0} P_\auf (1, \x) - \tst{\2}
	    = \tst{\2} \PH (\i \p|1) \epp
\end{equation}
Using (\ref{mirep}), (\ref{mirepq}) in the last expression on the
right hand side we obtain \cite{GKS04a}
\begin{equation} \label{magnet}
     m(T,h) = - \2 - \int_{\cal C} \frac{d \om}{2 \p \i}
                   \frac{G(\om,0)}{1 + \fa (\om)}
            = \2 + \int_{\cal C} \frac{d \om}{2 \p \i}
                   \frac{G(\om,0)}{1 + \faq (\om)} \epp
\end{equation}

In the following sections we shall consider the limits $\h \rightarrow
\i \p/2$ and $\h \rightarrow \infty$ with $J \D = c$ fixed. In the
former case $\D$ goes to zero and the Hamiltonian (\ref{xxzham})
becomes the XX Hamiltonian. This limit is also called the free Fermion
limit. The latter limit is called the Ising limit, since in this
limit the Hamiltonian (\ref{xxzham}) turns into the Hamiltonian of
the one-dimensional Ising model.

\section{The XX limit}
In the XX limit $\h = \i \p/2$, $\D = \ch(\h) = 0$ we have
\begin{equation} \label{xxsh}
     \sh (\la \pm \h) = \pm \i \ch (\la) \epc
\end{equation}
and the equations (\ref{nlieh})-(\ref{pinintrep}) simplify
considerably. In particular, the kernel of the integral equations
(\ref{nlieh}) and (\ref{repdefg}) vanishes identically, and the
auxiliary function $\fa$ and the density function $G$ are given by
the corresponding inhomogeneities,
\begin{align} \label{xxfa}
     \fa (\la) & = \exp \Bigl\{
        - \frac{1}{T} \Bigl(h + \frac{4 \i J}{\sh(2 \la)} \Bigr)
	  \Bigr\} \epc \\ \label{xxg}
     G(\la, \z) & = - \frac{2}{\sh(2(\la - \z))} \epp
\end{align}
Note that the function $G(\la, \z)$ does not depend on the temperature
and therefore agrees with its zero temperature limit. For this
fact the following calculations bear many similarities with the
zero temperature case studied in \cite{KMST02c}.

\subsection*{Magnetization}
We wish to transform all integrals over ${\cal C}$ into integrals
over the `Brillouin zone' $[-\p, \p]$. To begin with a simple example,
we consider the equation (\ref{magnet}) for the magnetization.
Inserting (\ref{xxfa}) and (\ref{xxg}) we obtain
\begin{equation} \label{magnetint1}
     m(T, h) = \2 - \int_{\cal C} \frac{d \om}{\p \i \sh(2 \om)}
               \frac{1}{1 + \exp \bigl( \frac{1}{T}
			( h + \frac{4 \i J}{\sh(2 \om)}) \bigr)} \epp
\end{equation}
The contour of integration ${\cal C}$ consists of a part ${\cal C_+}$
above the real axis and a part ${\cal C_-}$ below the real axis (see
figure \ref{fig:cancon}). Since the integrand in (\ref{magnetint1})
is analytic in the strips $- \p/2 < \Im \om < 0$ and $0 < \Im \om <
\p/2$ we may deform ${\cal C_\pm}$ into contours parallel to the
real axis passing through the points $\pm \i \p/4$. Then the
parameterization
\begin{subequations}
\label{paracont}
\begin{align}
     & {\cal C_+}: \bigl[ \tst{\frac{\p}{2}}, \p \bigr] \cup
                   \bigl[ - \p, - \tst{\frac{\p}{2}} \bigr]
		   \rightarrow {\mathbb C}, &
		 \om (p) = \2 \arth(\sin(p)) + \frac{\i \p}{4} & \epc\\
     & {\cal C_-}: \bigl[ - \tst{\frac{\p}{2}},
                            \tst{\frac{\p}{2}} \bigr]
                 \rightarrow {\mathbb C}, &
		 \om (p) = \2 \arth(\sin(p)) - \frac{\i \p}{4} &
\end{align}
\end{subequations}
of ${\cal C}$ is uniform, and for every point $\om$ on $\cal C$ there
is a unique $p \in [- \p,\p]$ such that
\begin{equation} \label{unishch}
     \sh(2 \om) = \frac{1}{\i \cos(p)} \epc \qd
     \ch(2 \om) = - \i \tg(p)
\end{equation}
It follows the familiar expression
\begin{equation}
     m(T, h) = \2 - \int_{-\p}^\p \frac{d p}{2 \p}
               \frac{1}{1 + \exp \bigl( \frac{h - 4 J \cos p}{T}
	                                              \bigr)} \epp
\end{equation}

\subsection*{Emptiness formation probability}
The formula for the emptiness formation probability is obtained in a
similar way. We first insert (\ref{xxsh})-(\ref{xxg}) into
(\ref{pinintrep}) and use the product formula for the Cauchy
determinant. Then
\begin{equation}
     P_\auf (m|\{\x\}) = \frac{1}{m!}     
	\biggl[ \prod_{j=1}^m
	        \int_{\cal C} \frac{d \om_j}{2 \p \i (1 + \fa (\om_j))}
	\biggr] \\[1ex]
        \frac{\prod_{\substack{j,k = 1 \\ j \ne k}}
	      \sh(\om_j - \om_k) \ch(\x_j - \x_k)}
	     {\prod_{j,k = 1}^m \sh(\om_j - \x_k) \ch(\om_j - \x_k)}
	     \epc
\end{equation}
and the homogeneous limit $\x_j \rightarrow 0$ is obvious,
\begin{equation} \label{pxxc}
     P_\auf (m) = \frac{(-1)^{\frac{m(m-1)}{2}}}{m!}     
	\biggl[ \prod_{j=1}^m
	        \int_{\cal C} \frac{d \om_j}{\p \i (1 + \fa (\om_j))}
	\biggr] \\[1ex]
        \frac{\prod_{1 \le j < k \le m} 4 \sh^2 (\om_j - \om_k)}
	     {\prod_{j = 1}^m \sh^m (2 \om_j)} \epp
\end{equation}
Again the contour and the integrand are uniformly parameterized
by (\ref{paracont}), in particular, using (\ref{unishch}),
\begin{equation} \label{unishsq}
     \sh^2 (\om_j - \om_k) = \2 \ch(2 \om_j) \ch(2 \om_k)
        - \2 \sh(2 \om_j) \sh(2 \om_k) - \2 =
	\frac{\sin^2 \bigl( \frac{p_j - p_k}{2} \bigr)}
	     {\cos(p_j) \cos(p_k)}
\end{equation}
for {\it all} $p_j, p_k \in [- \p. \p]$. Inserting (\ref{paracont}),
(\ref{unishch}) and (\ref{unishsq}) into (\ref{pxxc}) we end up with
\begin{equation} \label{pxxbril}
     P_\auf (m) = \frac{1}{m!}     
	\Biggl[ \prod_{j=1}^m \int_{-\p}^\p \frac{d p_j}{2 \p}
	                      \frac{1}{1 + \exp \Bigl( \frac{
			      4 J \cos p_j - h}{T} \Bigr)}
	\Biggr] \\[1ex]
        \prod_{1 \le j < k \le m} 4
	       \sin^2 \Bigl( \frac{p_j - p_k}{2} \Bigr) \epp
\end{equation}
This can be further simplified, if we take into account that
$\prod_{1 \le j < k \le m} 4 \sin^2 \Bigl( \frac{p_j - p_k}{2} \Bigr)$
can be expressed as a product of two van-der-Monde determinants. Then
\begin{equation}
     P_\auf (m) = \det_m \Biggl[ \int_{-\p}^\p \frac{d p}{2 \p}
        \frac{\re^{\i (j - k)p}}
	     {1 + \exp \Bigl( \frac{ 4 J \cos p - h}{T} \Bigr)}
	     \Biggr]_{j, k = 1, \dots, m}
\end{equation}
which is the known result, first obtained in \cite{STN01}. For a
careful discussion of the large distance asymptotics of $P_\auf (m)$
at zero and non-zero temperature see also \cite{STN01}.

\subsection*{Two-point function}
For the calculation of the two-point function $\bigl\< S_1^z S_m^z
\bigr\>_{T, h}$ we use the integral representation (\ref{mirepq}).
Inserting (\ref{xxsh})-(\ref{xxg}) and applying the product formula for
the Cauchy determinant twice we find
\begin{multline} \label{xxgenfun1}
     \PH (\ph|m) = \\ \sum_{n=0}^m \frac{(\re^\ph - 1)^n}{(n!)^2}
	\biggl[ \prod_{j=1}^n \int_{\G} \frac{d \z_j}{2 \p \i}
	\int_{\cal C} \frac{d \om_j}{2 \p \i (1 + \faq (\om_j))}
        \biggl( \frac{\tgh(\om_j)}{\tgh(\z_j)} \biggr)^m \biggr]
	\biggl[ \det_n \frac{1}{\sh(\om_j - \z_k)} \biggr]^2 \epc
\end{multline}
where $\faq = 1/\fa$ is given explicitly in (\ref{xxfa}). The right
hand side has a rather similar structure as in the zero temperature
limit, and, at first instance, the same tricks apply (cf.\
\cite{KMST02c}). Since the parameter $\ph$ enters (\ref{xxgenfun1})
in the combination $(\re^\ph - 1)^m$, only the three lowest order
terms in the sum remain after applying the differential-difference
operator at the right hand side of (\ref{genfunappl}), and
\begin{equation} \label{xxtwo1}
     \bigl\< S_1^z S_m^z \bigr\>_{T, h} = \tst{\2} D_m^2 (I_1 + I_2)
        - D_m I_1 + \tst{\4} \epc
\end{equation}
where
\begin{subequations}
\label{i12}
\begin{align} \label{i1}
     I_1 & = \int_{\G} \frac{d \z}{2 \p \i}
	\int_{\cal C} \frac{d \om}{2 \p \i (1 + \faq (\om))}
        \biggl( \frac{\tgh(\om)}{\tgh(\z)} \biggr)^m
	\frac{1}{\sh^2 (\om - \z)} \epc \\[1ex] \label{i2}
     I_2 & = \2 \biggl[ \prod_{j=1}^2 \int_{\G} \frac{d \z_j}{2 \p \i}
	\int_{\cal C} \frac{d \om_j}{2 \p \i (1 + \faq (\om_j))}
        \biggl( \frac{\tgh(\om_j)}{\tgh(\z_j)} \biggr)^m \biggr]
	\biggl[ \det_2 \frac{1}{\sh(\om_j - \z_k)} \biggr]^2
	\epp
\end{align}
\end{subequations}

\begin{figure}

\begin{center}

\epsfxsize \textwidth
\epsffile{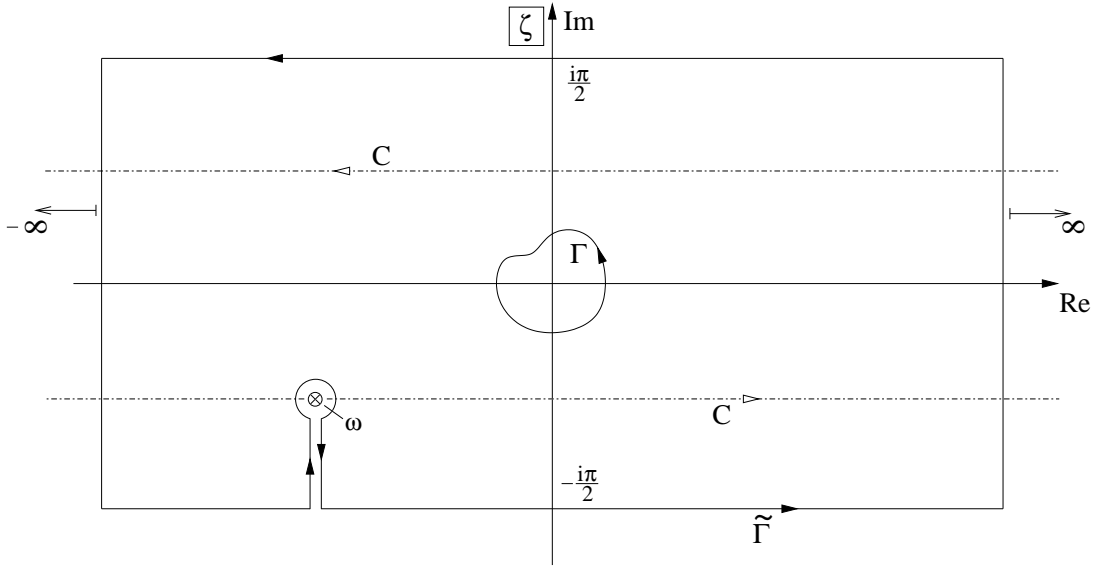}

\caption{\label{fig:xxcont} In order to calculate the integrals over
$\z$ in $I_1$ and $I_2$ the contour $\G$ is deformed into $\tilde \G$
plus a small circle surrounding the pole at $\om$.
}
\end{center}
\end{figure}

The $\G$-integrals in (\ref{i12}) can be calculated by means of residue
calculus. For this purpose we deform the contour $\G$ as sketched
in figure \ref{fig:xxcont}. Then
\begin{multline} \label{calci1}
     \int_\G \frac{d \z}{2\p \i}
        \biggl( \frac{\tgh(\om)}{\tgh(\z)} \biggr)^m
	\frac{1}{\sh^2 (\om - \z)} =
	   \int_{\tilde \G} \frac{d \z}{2\p \i}
	   \biggl( \frac{\tgh(\om)}{\tgh(\z)} \biggr)^m
	   \frac{1}{\sh^2 (\om - \z)} \\[1ex]
	   - \res \biggl\{
	   \biggl( \frac{\tgh(\om)}{\tgh(\z)} \biggr)^m
	   \frac{1}{\sh^2 (\om - \z)} \biggr\}_{\z = \om}
	   = \frac{2m}{\sh (2 \om)} \epc
\end{multline}
where the integral over $\tilde \G$ is zero, because the integrand is
periodic with period $\i \p$ and vanishes asymptotically for
$\Re \z \rightarrow \pm \infty$. Inserting (\ref{calci1}) into
(\ref{i1}) and comparing with (\ref{magnetint1}) we conclude that
\begin{equation}
     I_1 = m \bigl[ \tst{\2} - m(T, h) \bigr]
         = m \bigl[ \tst{\2} - \bigl\< S_j^z \bigr\>_{T, h} \bigr] \epp
\end{equation}
Similarly
\begin{multline}
     I_2 = I_1^2
         - \biggl[ \prod_{j=1}^2
	   \int_{\cal C} \frac{d \om_j \, \tgh^m (\om_j)}
	                      {2 \p \i (1 + \faq (\om_j))} \biggr]
           \biggl[ \int_{\G} \frac{d \z}{2 \p \i}
	           \frac{1}
		        {\tgh^m (\z) \sh(\z - \om_1) \sh(\z - \om_2)}
			 \biggr]^2 \\
           = I_1^2
           - \biggl[ \prod_{j=1}^2
	     \int_{\cal C} \frac{d \om_j}{2 \p \i (1 + \faq (\om_j))}
	                   \biggr]
             \biggl[ \frac{1}{\sh(\om_1 - \om_2)}
	             \biggl( \biggl[ \frac{\tgh (\om_1)}{\tgh(\om_2)}
		             \biggr]^\frac{m}{2} -
	                     \biggl[ \frac{\tgh (\om_2)}{\tgh(\om_1)}
		             \biggr]^\frac{m}{2} \biggr) \biggr]^2 \epp
\end{multline}
Inserting the above expressions for $I_1$ and $I_2$ into (\ref{xxtwo1})
and applying the difference operators we arrive at
\begin{multline}
     \bigl< S_1^z S_{m+1}^z \bigr\>_{T, h}
        - \bigl< S_1^z \bigr\>_{T, h} \bigl< S_{m+1}^z \bigr\>_{T, h} 
	= \\[1ex] \biggl[
	     \int_{\cal C} \frac{d \om}{2 \p \i (1 + \faq (\om))}
	     \frac{\6_\om \tgh^m (\om)}{m} \biggr] \biggl[
	     \int_{\cal C} \frac{d \om}{2 \p \i (1 + \faq (\om))}
	     \frac{\6_\om \cth^m (\om)}{m} \biggr] \epp
\end{multline}
This is already a rather simple expression involving only products of
single integrals. We wrote it in a form that is particularly
convenient for transforming it into an integral over the Brillouin
zone: Since (see (\ref{unishch}))
\begin{equation}
     \tgh (\om) = \frac{1}{\cth (\om)}
                = \frac{\ch (2 \om) - 1}{\sh (2 \om)}
		= - \i \re^{\i p} \epc
\end{equation}
we see at once that
\begin{equation} \label{xxtwop}
     \bigl< S_1^z S_{m+1}^z \bigr\>_{T, h}
        - \bigl< S_1^z \bigr\>_{T, h} \bigl< S_{m+1}^z \bigr\>_{T, h} 
	= - \Biggl[ \int_{-\p}^\p \frac{d p}{2 \p}
	    \frac{\cos(mp)}
	         {1 + \exp \bigl( \frac{h - 4 J \cos p}{T} \bigr)}
		 \Biggr]^2
\end{equation}
for all $m \in {\mathbb N}$.

Let us finally consider the zero temperature limit. If $h$ is larger
than the critical field $4J$, all spins are ferromagnetically aligned
at zero temperature and the correlation function (\ref{xxtwop})
vanishes. For $|h| < 4J$ the integral (\ref{xxtwop}) in the
zero temperature limit is restricted to the interval between two
Fermi points $\pm \arccos \bigl( \frac{h}{4J} \bigr)$, where the
exponential function in the integrand vanishes, and
\begin{equation} \label{xxtwopzero}
     \bigl< S_1^z S_{m+1}^z \bigr\>_{T, h}
        - \bigl< S_1^z \bigr\>_{T, h} \bigl< S_{m+1}^z \bigr\>_{T, h} 
	= \frac{T_{2m} \bigl( \frac{h}{4J} \bigr) - 1}{2 \p^2 m^2} \epc
\end{equation}
with $T_j (x)$ being the $j$th Chebychev polynomial. Equations
(\ref{xxtwop}) and (\ref{xxtwopzero}) are equivalent to formulae
first obtained in \cite{LSM61} (for $h = 0$) and \cite{BaMc71}.
The asymptotics of the two-point functions was studied in
\cite{BaMc71}.

\section{The Ising limit}
The XXZ Hamiltonian (\ref{xxzham}) turns into the Hamiltonian of the
one-dimensional Ising model if we set $J = \frac{c}{\D} =
\frac{c}{\ch (\h)}$, $c > 0$, and take the limit $\h \rightarrow
\infty$. In this section we shall reproduce the known formula for the
emptiness formation probability in the Ising limit \cite{Suzuki02} from
the multiple integral representation (\ref{pinintrep}). As a
prerequisite we need to calculate the auxiliary function $\fa (\la)$
and the inhomogeneous density function $G (\la, \z)$ from their
integral equations (\ref{nlieh}) and (\ref{repdefg}). We recall that,
in the off-critical regime, the integral equations determine the
functions $\fa(\la)$ and $G(\la, \z)$ inside the strip $|\Im \la| \le
\frac{\p}{2}$, $|\Re \la| \le (1 - \de) \frac{\h}{2}$ for any fixed
$\de$ with $0 < \de < 1$. Using that
\begin{equation} \label{limcth}
     \lim_{\h \rightarrow \infty} \cth(\la \pm \h) = \pm 1 \epc
\end{equation}
uniformly for all $\la$ with $|\Re \la| \le (1 - \de) \h$, and the
elementary formulae
\begin{align}
     \frac{\sh(\h)}{\sh(\la) \sh(\la + \h)} &
        = \cth(\la) - \cth(\la + \h) \epc \\
     \frac{\sh(2\h)}{\sh(\la - \h) \sh(\la + \h)} &
        = \cth(\la - \h) - \cth(\la + \h) \epc \label{addcth2}
\end{align}
we see that the integral equations (\ref{nlieh}), (\ref{repdefg})
turn into
\begin{align} \label{nlieaisi}
     \ln \fa (\la) & = - \frac{h}{T}
                       - \frac{2c}{T} \bigl( \cth(\la) - 1 \bigr)
                       + \int_{\cal C} \frac{d \om}{\p \i} \,
		         \ln (1 + \fa (\om)) \epc \\ \label{gisi}
     G (\la, \z) & = \cth(\z - \la) - 1
                     - \int_{\cal C} \frac{d \om}{\p \i} \,
		       \frac{G(\om, \z)}{1 + \fa (\om)} \epp
\end{align}
These equations are now valid for all $\la \in {\mathbb C}$ with
$|\Im \la| \le \frac{\p}{2}$. In particular, we may choose for $\la$
any point on the left or right edge of the contour $\cal C$ (see figure
\ref{fig:cancon}) and then shift the left edge to $- \infty$ and
the right edge to $+ \infty$. This way we obtain from both integral
equations, (\ref{nlieaisi}) and (\ref{gisi}), two algebraic
equations that determine the limits $\fa_\pm = \lim_{|\Re \la|
\rightarrow \pm \infty} \fa (\la)$ and $G_\pm = \lim_{|\Re \la|
\rightarrow \pm \infty} G (\la, \z)$, respectively. The equations
for the limits of the auxiliary functions read
\begin{equation} \label{apm}
     \fa_\pm = K \re^{- \frac{h}{T} \mp \frac{2c}{T}} \epc \qd
     K = \frac{1 + \fa_+}{1 + \fa_-} \, \re^{\frac{2c}{T}} \epp
\end{equation}
This is easily turned into a quadratic equation for $K$ whose roots are
\begin{equation} \label{kroot}
     K = \re^{- \frac{2c}{T} + \frac{h}{2T}}
         \Biggl( - \sh \Bigl( \frac{h}{2T} \Bigr) \pm
         \sqrt{\sh^2 \Bigl( \frac{h}{2T} \Bigr) + \re^{\frac{4c}{T}}}
	    \Biggr) \epp
\end{equation}
Shifting the left and right integration contours to $\mp \infty$
the constant $K$ appears in (\ref{nlieaisi}), and we
end up with
\begin{equation} \label{auxisi}
     \fa(\la) = K \re^{- \frac{h}{T} -\frac{2c}{T} \cth(\la)} \epp
\end{equation}
The sign in (\ref{kroot}) is fixed to plus by taking into account that
$\fa (\la)$ in the high temperature limit is identically equal to 1
for arbitrary value of the anisotropy parameter $\D$ (see
\cite{GKS04a}).

The function $G(\la, \z)$ is obtained in a similar way. The limits
$G_\pm$ satisfy
\begin{equation} \label{gpmeq}
     G_+ = - 2 - \frac{G_+}{1 + \fa_+} + \frac{G_-}{1 + \fa_-} \epc \qd
     G_- = - \frac{G_+}{1 + \fa_+} + \frac{G_-}{1 + \fa_-}
\end{equation}
which follows from (\ref{gisi}). We infer that $G_\pm$ are independent
of $\z$ and that $G_- - G_+ = 2$. Hence we can introduce a new
function, say $\<\s\>$, such that
\begin{equation} \label{defsigma}
     G_\pm = \<\s\> \mp 1 \epp
\end{equation}
Inserting the latter definition into (\ref{gpmeq}), using (\ref{apm}),
(\ref{kroot}) and solving for $\<\s\>$, we obtain the explicit
expression
\begin{equation} \label{sigmaexp}
     \<\s\> = \frac{\sh \bigl( \frac{h}{2T} \bigr)}
                   {\sqrt{\sh^2 \bigl( \frac{h}{2T} \bigr) +
		    \re^{\frac{4c}{T}}}} \epp
\end{equation}
By shifting the integration contour to $\mp \infty$ for fixed $\la$
in (\ref{gisi}) and using (\ref{gpmeq}), (\ref{defsigma}) we eventually
arrive at
\begin{equation}
     G(\la, \z) = \cth (\z - \la) + \<\s\> \epp
\end{equation}

Having determined the auxiliary function $\fa (\la)$ and the function
$G(\la, \z)$ we can now calculate the magnetization as a first
simple application. Shifting the integration contour to $\mp \infty$
in the first equation (\ref{magnet}) and making use of (\ref{gpmeq})
and (\ref{defsigma}) we obtain at once
\begin{equation} \label{magisi}
     m(T, h) = \frac{\<\s\>}{2} \epc
\end{equation}
i.e.\ $\<\s\> = \<\s_j^z\>_{T, h}$ which explains our notation. Not
totally unexpected  our formulae (\ref{magisi}) and (\ref{sigmaexp})
are in agreement with the well-known result \cite{Babook}.

\subsection*{Emptiness formation probability}
\begin{figure}
\begin{center}
\epsfxsize \textwidth
\epsffile{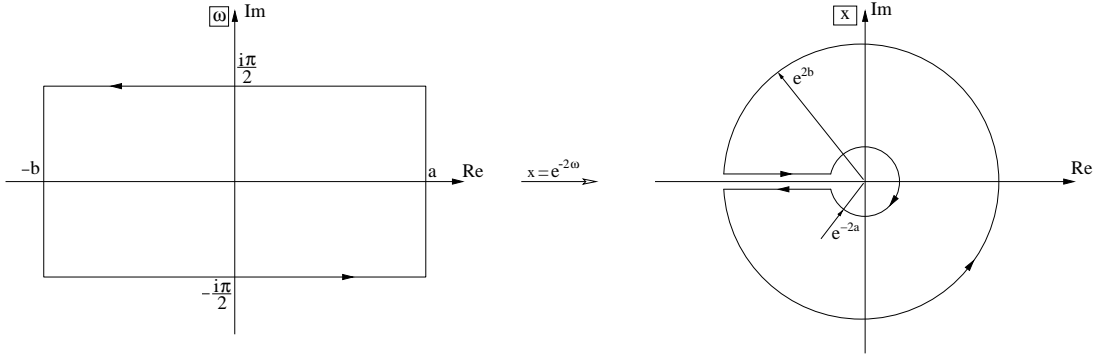}

\caption{\label{fig:transform}
The transformation of the canonical contour under the map
$\re^{-2 \om} \rightarrow x$.
}
\end{center}
\end{figure}

Starting point for the calculation of the emptiness formation
probability is again (\ref{pinintrep}). After (i) taking into account
the simplifications due to (\ref{limcth})-(\ref{addcth2}), (ii) using
the formula
\begin{equation}
     \det_m \frac{a x_j + b s_k}{x_j - s_k} = (a + b)^{m-1}
        \biggl[ a \prod_{k=1}^m x_k + b \prod_{k=1}^m s_k \bigg]
	\frac{\prod_{1 \le j < k \le m} (x_j - x_k)(s_k - s_j)}
	     {\prod_{j,k = 1}^m (x_j - s_k)}
\end{equation}
in order to simplify $\det_m G(\om_j,\x_k)$, and (iii) setting
$s_k = \re^{- 2 \x_k}$ and changing the integration variables to
$x_j = \re^{- 2 \om_j}$, we arrive at
\begin{multline} \label{emptisi1}
     P_\auf (m|\{\x\}) = \frac{1}{2m!}
        \biggl[ \prod_{j=1}^m \int_{\cal T}
	        \frac{d x_j}{2 \p \i \, x_j}
		\frac{1}{1 + K \exp \Bigl(- \frac{h}{T} - \frac{2c}{T}
		                    \frac{1 + x_j}{1 - x_j} \Bigr)}
			            \biggr] \\
        \biggl[ \prod_{\substack{j,k = 1\\j \ne k}}^m (x_j - x_k)
	        \biggr]
        \frac{(1 + \<\s\>) \prod_{k=1}^m x_k
	         + (1 - \<\s\>) \prod_{k=1}^m s_k}
             {\prod_{j,k = 1}^m (x_j - s_k)} \epp
\end{multline}
Here we denoted the transformed contour by ${\cal T}$. It consists of
two concentric circles about the origin. The left edge at, say, $-b$
of the original contour of integration is mapped to a large circle of
radius $\re^{2b}$, while the right edge at, say, $a$ is mapped to a
small circle of radius $\re^{-2a}$ (see figure \ref{fig:transform}).
As the radius of the large circle is increased the auxiliary function
on the contour tends to its limiting value $\fa_- = K
\re^{-(h - 2c)/T}$. On the small circle it approaches $\fa_+ = K
\re^{- (h + 2c)/T}$ if the radius goes to zero. Since the remaining
factors in the integrand are rational in the $x_j$ with only simple
poles at $x_j = s_k$, $k = 1, \dots, m$, and at $x_j = 0$, we may
calculate the integral by first fixing a large and a small circle,
such that all poles of the rational part of the integrand, except the
poles at zero, lie in the annulus between the circles, then replacing
the auxiliary function by $\fa_-$ on the large circle and $\fa_+$ on
the small circle and finally using the residue theorem. After carrying
out $n$ integration this way the right hand side of (\ref{emptisi1})
takes the form
\begin{align}
     P_\auf (m|\{\x\}) & = \frac{1}{2m!(1 + \fa_-)^n}
        \biggl[ \prod_{j=1}^{m - n} \int_{\cal T}
	        \frac{d x_j}{2 \p \i \, x_j}
		\frac{1}{1 + K \exp \Bigl(- \frac{h}{T} - \frac{2c}{T}
		                    \frac{1 + x_j}{1 - x_j} \Bigr)}
			            \biggr]
        \biggl[ \prod_{\substack{j,k = 1\\j \ne k}}^{m - n} (x_j - x_k)
	        \biggr] \notag \\[-1ex]
        \sum_{j_1 = 1}^m \sum_{\substack{j_2 = 1\\j_2 \ne j_1}}^m &
	\dots \sum_{\substack{j_{n-1} = 1\\
	                      j_{n-1} \ne j_1, \dots, j_{n-2}}}^m
	\Biggl\{
	\sum_{\substack{j_n = 1\\j_n \ne j_1, \dots, j_{n-1}}}^m
	\frac{\prod_{k=1}^{m-n} \prod_{\ell = 1}^n (s_{j_\ell} - x_k)}
	     {\prod_{\substack{k = 1\\k \ne j_1, \dots, j_n}}^m
	      \prod_{\ell = 1}^n (s_{j_\ell} - s_k)} \notag \\
        & \mspace{180.mu}
	\frac{(1 + \<\s\>) \prod_{k=1}^{m - n} x_k
	         + (1 - \<\s\>)
		   \prod_{\substack{k = 1\\k \ne j_1, \dots, j_n}}^m
		          s_k}
             {\prod_{j = 1}^{m - n}
	      \prod_{\substack{k = 1\\k \ne j_1, \dots, j_n}}^m 
	             (x_j - s_k)} \notag \\[1ex]
        - n (1 - & \<\s\>) \frac{\fa_+ - \fa_-}{1 + \fa_+}
	\frac{\Bigl[ \prod_{k=1}^{m-n} \prod_{\ell = 1}^{n - 1}
	      (s_{j_\ell} - x_k) \Bigr] \Bigl[
	      \prod_{\ell = 1}^{n-1} s_{j_\ell} \Bigr]}
	     {\prod_{\substack{k = 1\\k \ne j_1, \dots, j_{n-1}}}^m
	      \prod_{\ell = 1}^{n-1} (s_{j_\ell} - s_k)}
	\frac{\prod_{k=1}^{m - n} x_k^2}
             {\prod_{j = 1}^{m - n}
	      \prod_{\substack{k = 1\\k \ne j_1, \dots, j_{n-1}}}^m 
	             \mspace{-18.mu} (x_j - s_k)} \Biggr\}
\end{align}
which is most easily verified by induction over $n$.

For $n = m$ numerous trivial cancellations among the products occur,
and we obtain
\begin{equation} \label{almostisi}
     P_\auf (m|\{\x\}) = \frac{1}{(1 + \fa_-)^m}
        \biggl[ 1 - \frac{1 - \<\s\>}{2} \,
	            \frac{\fa_+ - \fa_-}{1 + \fa_+}
		    \sum_{k=1}^m
		    \prod_{\substack{\ell = 1\\ \ell \ne k}}^m
		    \frac{s_\ell}{s_\ell - s_k} \biggr] \epp
\end{equation}
This further simplifies due to the identity
\begin{equation} \label{sumid}
     \sum_{k=1}^m \prod_{\substack{\ell = 1\\ \ell \ne k}}^m
        \frac{s_\ell}{s_\ell - s_k} = 1
\end{equation}
which holds for any set of mutually distinct complex numbers $s_k$,
$k = 1, \dots, m$. Inserting (\ref{sumid}) and (\ref{gpmeq}),
(\ref{defsigma}) into (\ref{almostisi}) we end up with
\begin{equation} \label{pisi1}
     P_\auf (m|\{\x\}) = \frac{1 + \<\s\>}{2 (1 + \fa_-)^{m-1}} \epp
\end{equation}
where $\<\s\>$ and $\fa_-$ are explicitly shown in equations
(\ref{sigmaexp}) and (\ref{apm}), (\ref{kroot}).  Quite remarkably,
the right hand side of (\ref{pisi1}) is completely independent of the
inhomogeneities $\{\x\}$.

In order to express the emptiness formation probability in more
physical terms let us calculate the free energy in the Ising limit
from the integral (\ref{freee}). Proceeding as above we find
\begin{align} \label{freeisi}
     f (T, h) & = -\frac{h}{2} - T \ln(1 + \fa_-) \notag \\
              & = - T \ln \Biggl( \ch \Bigl( \frac{h}{2T} \Bigr)
	            + \sqrt{\sh^2 \Bigl( \frac{h}{2T} \Bigr)
		            + \re^\frac{4c}{T}} \Biggr) \epp
\end{align}
Comparison of (\ref{pisi1}) with (\ref{magisi}) and (\ref{freeisi})
then yields
\begin{equation}
     P_\auf (n + 1) = \bigl( \tst{\2} + m(T, h) \bigr)
                      \re^{\frac{n f(T, h)}{T} + \frac{n h}{2T}}
\end{equation}
in accordance with \cite{Suzuki02} and with the general formula
for the asymptotics of the emptiness formation probability at finite
temperature \cite{BoKo01}.

\section{Conclusions}
Our aim in this article was to demonstrate the usefulness of finite
temperature multiple integral representations for correlation
functions, as obtained in \cite{GKS04a,GKS04bpp,GKS05}, by showing
that a number of known results for the correlation functions of the
XX chain and of the one-dimensional Ising model can be reproduced by
means of our formulae (\ref{mirep}), (\ref{mirepq}) and
(\ref{pinintrep}). Usually the correlation functions of the XX and
of the Ising chain are calculated from rather different, very special
devices. The correlation functions of the XX chain are obtained
by mapping the chain onto a model of free Fermions \cite{LSM61},
while the calculation of the correlation functions of the
one-dimensional Ising model relies on the transfer matrix formalism
with a spectral parameter independent $2 \times 2$ transfer matrix
\cite{Babook}. Here, for the first time, we have obtained correlation
functions for the XX chain and for the Ising chain from a common
device, namely the multiple integral representation, which interpolates
for all values of $\D$ between the XX limit $\D = 0$ and the Ising
limit $\D \rightarrow \infty$.

We would like to see this work in a line with \cite{TsSh05}, where a
high-order high-temperature expansion for correlation functions of
the isotropic chain was obtained starting from the multiple integrals,
and with \cite{BoGo05}, where, again from the multiple integrals, some
of the two- and three-site functions were computed numerically. We hope
to report on further applications of the finite temperature integral
formulae in the future.

{\bf Acknowledgement.}
The authors would like to thank H. Boos and A. Kl\"umper for helpful
discussions. This work was partially supported by the Deutsche
Forschungsgemeinschaft under grant number Go 825/4-2.


\begin{thebibliography}{10}

\bibitem{BaMc71}
E.~Barouch and B.~M. McCoy, \emph{Statistical mechanics of the {XY} model.
  {II}. {S}pin correlation functions}, Phys. Rev. A \textbf{3} (1971), 786.

\bibitem{Babook}
R.~J. Baxter, \emph{Exactly solved models in statistical mechanics}, Academic
  Press, London, 1982.

\bibitem{BJMST04a}
H.~Boos, M.~Jimbo, T.~Miwa, F.~Smirnov, and Y.~Takeyama, \emph{A recursion
  formula for the correlation functions of an inhomogeneous {XXX} model},
  preprint, hep-th/0405044, 2004.

\bibitem{BJMST04b}
\bysame, \emph{Reduced $q${KZ} equation and correlation functions of the {XXZ}
  model}, preprint, hep-th/0412191, 2004.

\bibitem{BoKo01}
H.~E. Boos and V.~E. Korepin, \emph{Quantum spin chains and {R}iemann zeta
  function with odd arguments}, J. Phys. A \textbf{34} (2001), 5311.

\bibitem{BKS03}
H.~E. Boos, V.~E. Korepin, and F.~A. Smirnov, \emph{Emptiness formation
  probability and quantum {K}nizhnik-{Z}amolodchikov equation}, Nucl. Phys. B
  \textbf{658} (2003), 417.

\bibitem{BoGo05}
M.~Bortz and F.~G\"ohmann, \emph{Exact thermodynamic limit of short range
  correlation functions of the antiferromagnetic {XXZ} chain at finite
  temperatures}, preprint, cond-mat/0504370, 2005.

\bibitem{GKS04bpp}
F.~G\"ohmann, A.~Kl\"umper, and A.~Seel, \emph{Emptiness formation probability
  at finite temperature for the isotropic {H}eisenberg chain}, preprint,
  cond-mat/0406611, to be published in Physica B, 2004.

\bibitem{GKS04a}
\bysame, \emph{Integral representations for correlation functions of the {XXZ}
  chain at finite temperature}, J. Phys. A \textbf{37} (2004), 7625.

\bibitem{GKS05}
\bysame, \emph{Integral representation of the density matrix of the {XXZ} chain
  at finite temperature}, J. Phys. A \textbf{38} (2005), 1833.

\bibitem{InSu88}
M.~Inoue and M.~Suzuki, \emph{The {ST}-transformation approach to analytic
  solutions of quantum systems. {II}.\ {T}ransfer-matrix and {P}faffian
  methods}, Prog. Theor. Phys. \textbf{79} (1988), 645.

\bibitem{IzKo84}
A.~G. Izergin and V.~E. Korepin, \emph{The quantum inverse scattering method
  approach to correlation functions}, Comm. Math. Phys. \textbf{94} (1984), 67.

\bibitem{JMMN92}
M.~Jimbo, K.~Miki, T.~Miwa, and A.~Nakayashiki, \emph{Correlation functions of
  the {XXZ} model for {$\Delta < - 1$}}, Phys. Lett. A \textbf{168} (1992),
  256.

\bibitem{JiMi96}
M.~Jimbo and T.~Miwa, \emph{Quantum {KZ} equation with $|q| = 1$ and
  correlation functions of the {XXZ} model in the gapless regime}, J. Phys. A
  \textbf{29} (1996), 2923.

\bibitem{KSTS03}
G.~Kato, M.~Shiroishi, M.~Takahashi, and K.~Sakai, \emph{Next-nearest-neighbour
  correlation functions of the spin-1/2 {XXZ} chain at the critical region}, J.
  Phys. A \textbf{36} (2003), L337.

\bibitem{KSTS04}
\bysame, \emph{Third-neighbour and other four-point correlation functions of
  spin-1/2 {XXZ} chain}, J. Phys. A \textbf{37} (2004), 5097.

\bibitem{KMST02c}
N.~Kitanine, J.~M. Maillet, N.~A. Slavnov, and V.~Terras, \emph{Correlation
  functions of the {XXZ} spin-$\frac{1}{2}$ {H}eisenberg chain at the free
  fermion point from their multiple integral representations}, Nucl. Phys. B
  \textbf{642} (2002), 433.

\bibitem{KMST02a}
\bysame, \emph{Spin-spin correlation functions of the {XXZ}-$\frac{1}{2}$
  {H}eisenberg chain in a magnetic field}, Nucl. Phys. B \textbf{641} (2002),
  487.

\bibitem{KMST04bpp}
\bysame, \emph{Dynamical correlation functions of the {XXZ} spin-1/2 chain},
  preprint, hep-th/0407108, 2004.

\bibitem{KMST04cpp}
\bysame, \emph{On the spin-spin correlation functions of the {XXZ}
  spin-$\frac{1}{2}$ infinite chain}, preprint, hep-th/0407223, 2004.

\bibitem{KMT99b}
N.~Kitanine, J.~M. Maillet, and V.~Terras, \emph{Correlation functions of the
  {XXZ} {H}eisenberg spin-$\frac{1}{2}$ chain in a magnetic field}, Nucl. Phys.
  B \textbf{567} (2000), 554.

\bibitem{Kluemper92}
A.~Kl\"umper, \emph{Free energy and correlation length of quantum chains
  related to restricted solid-on-solid lattice models}, Ann.\ Physik \textbf{1}
  (1992), 540.

\bibitem{Kluemper93}
\bysame, \emph{Thermodynamics of the anisotropic spin-1/2 {H}eisenberg chain
  and related quantum chains}, Z. Phys. B \textbf{91} (1993), 507.

\bibitem{LSM61}
E.~H. Lieb, T.~Schultz, and D.~Mattis, \emph{Two soluble models of an
  antiferromagnetic chain}, Ann. Phys. (N.Y.) \textbf{16} (1961), 407.

\bibitem{STN01}
M.~Shiroishi, M.~Takahashi, and Y.~Nishiyama, \emph{Emptiness formation
  probability for the one-dimensional isotropic {XY} model}, J. Phys. Soc. Jpn.
  \textbf{70} (2001), 3535.

\bibitem{Suzuki85}
M.~Suzuki, \emph{Transfer-matrix method and {Monte Carlo} simulation in quantum
  spin systems}, Phys. Rev. B \textbf{31} (1985), 2957.

\bibitem{Suzuki02}
\bysame, \emph{Asymptotics and thermodynamics of spin pattern formation
  probability in equilibrium spin systems}, Phys. Lett. A \textbf{301} (2002),
  398.

\bibitem{TsSh05}
Z.~Tsuboi and M.~Shiroishi, \emph{High temperature expansion of emptiness
  formation probability for isotropic {H}eisenberg chain}, preprint,
  cond-mat/0502569, 2005.

\end{thebibliography}

\providecommand{\bysame}{\leavevmode\hbox to3em{\hrulefill}\thinspace}
\providecommand{\MR}{\relax\ifhmode\unskip\space\fi MR }
\providecommand{\MRhref}[2]{%
  \href{http://www.ams.org/mathscinet-getitem?mr=#1}{#2}
}
\providecommand{\href}[2]{#2}

\end{document}